\documentclass[sigconf]{acmart}

\AtBeginDocument{%
  }

\copyrightyear{2024}
\acmYear{2024}
\setcopyright{acmlicensed}\acmConference[CIKM '24]{Proceedings of the 33rd ACM International Conference on Information and Knowledge Management}{October 21--25, 2024}{Boise, ID, USA}
\acmBooktitle{Proceedings of the 33rd ACM International Conference on Information and Knowledge Management (CIKM '24), October 21--25, 2024, Boise, ID, USA}
\acmDOI{10.1145/3627673.3680063}
\acmISBN{979-8-4007-0436-9/24/10}

\begin{document}

\title{Advancing Re-Ranking with Multimodal Fusion and Target-Oriented Auxiliary Tasks in E-Commerce Search}

\author{Enqiang Xu}
\affiliation{%
  \institution{JD.com, Inc.}
  \city{Beijing}
  \country{China}
}
\email{xuenqiang@jd.com}
\orcid{0009-0002-4647-3439}

\author{Xinhui Li}
\affiliation{%
  \institution{JD.com, Inc.}
  \city{Beijing}
  \country{China}
}
\email{lixinhui9@jd.com}

\author{Zhigong Zhou}
\affiliation{%
  \institution{JD.com, Inc.}
  \city{Beijing}
  \country{China}
}
\email{zhouzhigong1@jd.com}

\author{Jiahao Ji}
 \authornote{Corresponding author}
\affiliation{%
  \institution{JD.com, Inc.}
  \city{Beijing}
  \country{China}
}
\email{jiahaoji@buaa.edu.cn}

\author{Jinyuan Zhao}
 \authornote{Corresponding author}
\affiliation{%
  \institution{JD.com, Inc.}
  \city{Beijing}
  \country{China}
}
\email{zhaojinyuan1@jd.com}

\author{Dadong Miao}
\affiliation{%
  \institution{JD.com, Inc.}
  \city{Beijing}
  \country{China}
}
\email{miaodadong@jd.com}

\author{Songlin Wang}
\affiliation{%
  \institution{JD.com, Inc.}
  \city{Beijing}
  \country{China}
}
\email{wangsonglin3@jd.com}

\author{Lin Liu}
\affiliation{%
  \institution{JD.com, Inc.}
  \city{Beijing}
  \country{China}
}
\email{liulin1@jd.com}

\author{Sulong Xu}
\affiliation{%
  \institution{JD.com, Inc.}
  \city{Beijing}
  \country{China}
}
\email{xusulong@jd.com}


\renewcommand{\shortauthors}{Enqiang Xu et al.}

\begin{abstract}
    In the rapidly evolving field of e-commerce, the effectiveness of search re-ranking models is crucial for enhancing user experience and driving conversion rates. Despite significant advancements in feature representation and model architecture, the integration of multimodal information remains underexplored. This study addresses this gap by investigating the computation and fusion of textual and visual information in the context of re-ranking. We propose \textbf{A}dvancing \textbf{R}e-Ranking with \textbf{M}ulti\textbf{m}odal Fusion and \textbf{T}arget-Oriented Auxiliary Tasks (ARMMT), which integrates an attention-based multimodal fusion technique and an auxiliary ranking-aligned task to enhance item representation and improve targeting capabilities. This method not only enriches the understanding of product attributes but also enables more precise and personalized recommendations. Experimental evaluations on JD.com's search platform demonstrate that ARMMT achieves state-of-the-art performance in multimodal information integration, evidenced by a 0.22\% increase in the Conversion Rate (CVR), significantly contributing to Gross Merchandise Volume (GMV). This pioneering approach has the potential to revolutionize e-commerce re-ranking, leading to elevated user satisfaction and business growth.
    
    % In the rapidly evolving field of e-commerce, the effectiveness of search ranking models is crucial for enhancing user experience and driving conversion rates. Despite the prevalent focus on optimizing feature representation and model architecture, the integration of multimodal information remains an underexplored area. This study fills this gap by investigating the computation and fusion of textual and visual data in the context of re-ranking. Drawing on the case of JD.com's search platform, we introduce an attention-based multimodal fusion technique and an auxiliary ranking-aligned task to enhance item representation and improve targeting capabilities. We propose a method called \textbf{A}dvancing \textbf{R}e-ranking with \textbf{M}ulti\textbf{m}odal Fusion and \textbf{T}arget-Oriented Auxiliary Tasks (ARMMT), which integrates these components to enhance re-ranking performance. It not only enriches the understanding of product attributes but also enables more precise and personalized recommendations. This pioneering approach holds the potential to revolutionize e-commerce re-ranking, leading to elevated user satisfaction and business growth. Experimental evaluations demonstrate that ARMMT achieves a new state-of-the-art in multimodal information integration, as evidenced by a 0.22\% increase in the User Conversion Rate (UCVR) on JD.com's platform, contributing significantly to the Gross Merchandise Volume (GMV).

\end{abstract}






\maketitle

\section{Introduction}
In the dynamic landscape of e-commerce, search ranking models are fundamental to enhancing user experience and optimizing conversion rates. These models play a crucial role in organizing the vast array of products that match a user's query, significantly impacting user satisfaction. Leading e-commerce platforms typically employ a cascade structure, which includes stages such as matching, pre-ranking, ranking, and re-ranking. The re-ranking stage, as the final step before a user's purchase decision, focuses on refining the scores of the top-k items by utilizing detailed user and item-specific information, thus providing a more personalized and relevant product list.

% In the vibrant landscape of e-commerce, search ranking models serve as the cornerstone of user experience and conversion optimization. These models are instrumental in organizing the array of products that respond to a user's query, a process that significantly influences user satisfaction and the probability of a transaction. The sorting pipeline within prominent e-commerce platforms typically adopts a multi-stage funnel approach, encompassing recall, pre-ranking, ranking, and re-ranking phases. The re-ranking stage, being the final touchpoint before user decision-making, is dedicated to refining the scoring of top-k items, leveraging user/item-specific information to deliver a more personalized and relevant product list.

Traditional re-ranking methods, often based on deep learning, primarily focus on model architecture and enhanced feature representation. These methods can be categorized into two types \cite{feng2021grn}. The first type is step-greedy re-ranking strategies \cite{zhuang2018globally,gong2022real,feng2021grn}, which determine the display result for each position sequentially, considering only the information of the previous item. This approach often falls short of the optimal outcome because it neglects the information of subsequent items. In contrast, contextual re-ranking strategies \cite{pei2019personalized,xi2021context,chen2022extr} capture the interdependencies among items using a contextual evaluation model, refining the click-through rate predictions for each item. For instance, PRM \cite{pei2019personalized} takes the initial ranking list as input and generates the optimal permutation based on the contextual model's predictions. Most existing re-ranking frameworks build upon this by adopting a two-stage architecture \cite{xi2021context,chen2022extr}, involving item sequence generation and ranking evaluation. The sequence generation task \cite{hidasi2018recurrent} refines the initial ranking results based on user queries and personalized information, while the ranking evaluation \cite{sakai2023versatile} ensures the quality of the generated results to meet user needs and provide a satisfactory user experience.

% Traditional re-ranking methods, often based on deep learning or graph neural networks, have primarily focused on model architecture and enhanced feature representation. Re-ranking model structures can be categorized into two types \cite{feng2021grn}. The first category is the step-greedy re-ranking strategies \cite{zhuang2018globally,gong2022real,feng2021grn}, which decide the display result for each position sequentially, considering only the information of the previous item and often falling short of the optimal outcome due to the neglect of subsequent items. In contrast, the second category is the contextual re-ranking strategies \cite{pei2019personalized,xi2021context,chen2022extr}, which capture the interdependencies among items using a contextual evaluation model and refines the click-through rate predictions for each item. The PRM \cite{pei2019personalized} approach, for instance, takes the initial ranking list as input and generates the optimal permutation based on the predictions from the contextual model. Building upon this, most existing re-ranking frameworks adopt a two-stage architecture \cite{xi2021context,chen2022extr}, involving item sequence generation and ranking evaluation. The sequence generation task \cite{hidasi2018recurrent} refines the preliminary sorting results based on user queries and personalized information while ranking evaluation \cite{sakai2023versatile} ensures the quality of the generated results to meet user needs and provide a satisfactory user experience.

These re-ranking models heavily rely on unique IDs and categorical features for user-item matching \cite{hidasi2018recurrent,zhou2019deep}. However, these methods primarily deal with sparse ID features and may not be adequately trained when the IDs appear infrequently in the data. In contrast, images provide intrinsic visual descriptions that can enhance model generalization. Given that users directly interact with item images, these images can offer additional visual information about user interests. Previous innovative works have introduced image features into recommendation systems \cite{cheng2016wide,mo2015image}, focusing on representing ads with image features in click-through rate prediction. For instance, AMS \cite{ge2018image} explored a new method based on a visual model to analyze user behavior using advanced model servers. However, most of these works concentrate on the acquisition of image features, with less research on the integration of multimodal features in ranking models. While progress has been made in extracting multimodal features \cite{radford2021learning}, there remains a gap in research on effectively incorporating these different types of features into re-ranking models.

% These re-ranking models heavily rely on unique IDs and categorical features for user-item matching \cite{hidasi2018recurrent,zhou2019deep}. However, these methods mainly deal with sparse ID features and may not be adequately trained when the IDs appear infrequently in the data. Images, on the other hand, provide intrinsic visual descriptions that can enhance model generalization. Considering that item images are directly interacted with by users, they can offer additional visual information about user interests. Previous creative works have introduced image information into recommendation systems \cite{cheng2016wide,mo2015image}, focusing on representing ads with image features in click-through rate prediction. AMS \cite{ge2018image} explored a new method based on a visual model to analyze user behavior using advanced model servers. However, most of these works concentrate on the acquisition of multimodal features, with less research on the integration of multimodal information in ranking models. While progress has been made in extracting multimodal features \cite{radford2021learning}, there remains a void in research on effectively incorporating such diverse data into ranking models.

This study bridges the gap by investigating the computation and fusion of multimodal information within the realm of re-ranking models, using JD.com's re-ranking system as a case study. The inclusion of multimodal cues, encompassing both textual and visual information, aims to mitigate the limitations of traditional textual and ID-based features, which lack the richness of visual information. To achieve this, we developed several key advancements. Firstly, we introduce an attention-based fusion mechanism, specifically the Context-Aware Fusion Unit (CAFU), which synergistically integrates textual and visual information. This method is designed to enhance item representation by incorporating visual information which might be neglected by traditional methods. Secondly, we propose the use of Multi-Perspective Self-Attention, which integrates multimodal information with other fields. This mechanism enhances the model's ability to capture complex interactions between different types of information, thereby improving the precision of the re-ranking process. Thirdly, to maximize the utility of multimodal information, we introduce an auxiliary task aligned with the ranking objective, providing additional supervision for the learning of multimodal representations. The integration of multimodal information into re-ranking models not only enhances our understanding of user preferences but also improves the precision and personalization of search results.

% This study bridges this gap by investigating the computation and fusion of multimodal data within the realm of re-ranking models, using JD.com's re-ranking system as a case study. The inclusion of multimodal cues, including textual and visual data, promises to mitigate the limitations of traditional textual and ID-based features, which lack the richness of visual content. The primary contributions of our research are as follows: Firstly, we introduce an attention-based fusion mechanism that synergistically integrates textual and visual information, representing an innovative strategy within the context of re-ranking models. This method is crafted to augment item representation by encapsulating visual attributes that are otherwise overlooked. Secondly, to maximize the utility of multimodal information, we propose an auxiliary task that aligns with the ranking objective, providing additional supervision for the learning of multimodal representations.

Our contributions can be summarized as follows:

\textbf{1. Attention-based Fusion Mechanism:} We introduce the Context-Aware Fusion Unit (CAFU) and Multi-Perspective Self-Attention, which integrate textual and visual information for re-ranking models. This approach addresses the limitations of traditional textual and ID-based features by incorporating rich visual cues, enhancing item representations.

\textbf{2. Auxiliary Task for Supervision:} We design an auxiliary task aligned with the ranking objective to supervise the learning process of multimodal representations, improving their quality and relevance for the ranking task.

\textbf{3. Empirical Validation:} We validate our approach through rigorous experimentation. Our results demonstrate the effectiveness of the CAFU, Multi-Perspective Self-Attention, and the auxiliary task in enhancing re-ranking models.

These contributions are significant for e-commerce. They introduce novel methods for optimizing re-ranking models, which have been successfully deployed, potentially improving user satisfaction, conversion rates, and the shopping experience.

% The integration of multimodal data into re-ranking models not only enhances our understanding of user preferences but also improves the precision and personalization of product recommendations. Our contributions can be summarized as follows:

% \textbf{Innovative Fusion Method}: We introduce an attention-based fusion method that combines textual and visual information for re-ranking models. This approach addresses the limitation of traditional textual features and ID-based features by incorporating rich visual information, which is essential for enhancing item representations.

% \textbf{Supervised Learning of Multimodal Representations}: To ensure the effectiveness of the multimodal information in the model, we design an auxiliary task that aligns with the ranking objective. This task supervises the learning process of the multimodal representations, improving their quality and relevance for the ranking task.

% \textbf{Empirical Validation}: Our research provides empirical validation of the proposed method through rigorous experimentation. The results are expected to demonstrate the effectiveness of the attention-based multimodal fusion approach in the context of re-ranking models.

% \textbf{Practical Implications}: The findings of this study have practical implications for the e-commerce industry, offering a new direction for optimizing re-ranking models. This can potentially lead to improved user satisfaction, increased conversion rates, and a better overall shopping experience.

\section{RELATED WORK}
In this section, we provide a concise overview of the latest developments in ranking models, with a focus on re-ranking algorithms and multimodal fusion within ranking frameworks. These areas are highly pertinent to the scope of our research.

\subsection{Re-ranking Models}
In the dynamic and competitive landscape of e-commerce, ranking models play a pivotal role in curating and presenting items in a manner that aligns with user preferences and needs. Traditional ranking models \cite{cheng2016wide, guo2017deepfm, lian2018xdeepfm, zhou2018deep, pi2020search, cao2022sampling, ma2018modeling} primarily focus on scoring individual items for click-through rate (CTR) estimation, aiming to optimize performance at a single point. Re-ranking models distinguish themselves from traditional ranking models by their unique ability to model the contextual relationships within a sequence of candidate items. These models are typically categorized into two distinct approaches: step-greedy strategies and context-wise strategies \cite{feng2021grn}.

\textbf{Step-greedy Re-ranking Strategies:} Step-greedy approaches utilize a sequential decision-making process for each position in the display results. These methods often employ recurrent neural networks or approximation solutions to determine the order of items. For instance, DPP \cite{chen2018fast} identifies the most relevant and diverse subset of candidates by calculating the determinant of a kernel matrix. In contrast, MMR \cite{carbonell1998use} relies on pairwise similarity. Seq2Slate \cite{bello2018seq2slate} uses a pointer network, while MIRNN \cite{ai2018learning} employs a gated recurrent unit to sequentially establish the order of items. However, this category of methods tends to overlook subsequent information in the ranking sequence, often resulting in suboptimal outcomes.

\textbf{Context-wise Re-ranking Strategies:} Context-wise methods aim to capture the mutual influence among items by using evaluation models that reassess the CTR or conversion rate (CVR) for each item. Methods like PRM \cite{pei2019personalized} and DLCM \cite{ai2018learning} process the initial ranking list with RNNs or self-attention mechanisms to model context signals and predict values for each item. To circumvent the evaluation-before-reranking dilemma \cite{xi2021context}, some researchers adopt a two-stage re-ranking framework, comprising permutation generation followed by permutation evaluation, such as GRN \cite{feng2021grn} and PRS \cite{feng2021revisit}. PIER \cite{shi2023pier} employs a fine-grained permutation selection module to choose the top-K candidates from the entire permutation space, along with a context-aware prediction module that predicts the list-wise CTR for each item.

In our approach, we adopt the two-stage architecture that emphasizes the seamless integration of multimodal information within a framework that captures reciprocal contextual interactions, thereby facilitating the identification of the most optimal sequence.

\subsection{Multimodal Fusion}
In the context of e-commerce recommendations, visual signals play a significant role as intuitive factors influencing user purchase decisions. Enhancing item representation with comprehensive visual information can enrich the description of the current item and elevate the model's level of personalization. Multimodal Recommender Systems \cite{liu2023multimodal} stand out from ID-based recommendation models due to their superior generalization capabilities \cite{yuan2023go}. Additionally, their proficiency in processing information across diverse modalities makes them particularly advantageous for multimedia services. Among the various feature fusion techniques, the attention mechanism is notably effective \cite{liu2023multimodal}, significantly enhancing the system's ability to understand and cater to user preferences. For instance, UVCAN \cite{liu2019user} utilizes user-side ID features to generate fusion weights for item-side multimodal information through self-attention, thereby facilitating micro-video recommendations.

In the realm of multimodal information fusion, several studies have made significant advancements. During its pre-training phase, MCPTR \cite{liu2022multi} leverages self-supervised multimodal contrastive learning to acquire fusion weights across different modalities, simultaneously deriving multimodal user and item representations. CMBF \cite{chen2021cmbf} adopts a cross-modal fusion approach to comprehensively integrate multimodal features, learning the interplay between different modalities. MML \cite{pan2022multimodal} and MARIO \cite{kim2022mario} employ an attention network to assess the impact of each modality on the interactions between users and items, preserving modality-specific attributes to obtain personalized embeddings for items relative to users. VLSNR \cite{han2022vlsnr} initially processes images and titles through a CLIP \cite{radford2021learning} encoder, followed by a series of attention layers to derive multimodal representations of news, and ultimately employs a GRU network to learn users' temporal interests.

Building on these studies, our approach focuses on the application of multimodal information fusion in ranking models. By integrating signals from different modalities, we aim to enhance the representation of items.

% Preliminaries和Method拆开，因为Method内容较多
\section{Preliminaries}
In this section, we initiate our discussion with the formal definitions of symbols that delineate the foundational framework of our re-ranking model and its associated tasks. Subsequently, we elucidate on the most prevalent model architectures in re-ranking, such as ID-based Deep Interest Network (DIN \cite{zhou2018deep}) and the context-based Transformer structure, which form the cornerstone of what we term as the base model.
% These are graphically represented on the upper right corner of Figure \ref{fig:framework}.
%这里，右上角图其实没有细化这部分

\subsection{Background} 
% 问题定义
In the architecture of conventional industrial search engine systems, a modular approach is typically employed, encompassing stages such as recall, pre-ranking, ranking, and re-ranking. Notably, the re-ranking phase serves to refine the output of the ranking module, optimizing the final presentation of search results to the end-user.

Mathematically, given a user $U$ entering a query $Q$ into the search bar, a ranking list of candidate items $I = \left \{ item_i \right \}_{i=1}^{N}$ is generated, where $N$ represents the top $N$ candidate items output by the ranking module. The task of the re-ranking stage is to learn a strategy $\mathcal{F}$ such that $I^{\ast} = \mathcal{F}(I, U, Q)$, which selects and rearranges items from $I$, presenting a final ranking list $I^{\ast}$ to users with the aim of maximizing the conversion rate (CVR).

\subsection{Base Model} 
\subsubsection{ID-based Deep Interest Network}
To capture user preferences from historical interactions, we utilize a target attention mechanism, specifically the Deep Interest Network (DIN). For each candidate item, the query is formed by concatenating the representations of the item ID, shop ID, and brand ID. Similarly, the key, representing the user's historical behavior sequence, is formed in the same way. By applying the attention mechanism, we derive the personalized item representation $P_\text{id}$ for each candidate item.

\subsubsection{Context-based Transformer Encoder}
We concatenate $P_\text{id}$ with item features to create the representation of candidate items. To capture the interactions among these candidate items, specifically as inspired by the Personalized Ranking Model (PRM), we linearly transform this representation to obtain $Q$, $K$, and $V$, and then utilize a transformer encoder to compute $A$, as shown in Equation \ref{3}.

\begin{equation}
    A = \text{Attention}(Q, K, V) = \text{softmax}(\frac{QK^\top}{\sqrt{d_k}})V
    \label{3}
\end{equation}
where, $d_k$ is the dimensionality of the key vectors $K$.

Subsequently, we pass $A$ through a fully connected layer to reduce its dimensionality. To effectively rank the candidate items, we leverage the listwise modeling capability by employing the softmax activation function, which outputs the predicted scores $\hat{y}$ for each item. We minimize the cross-entropy loss $\mathcal{L}$ between the conversion labels $y$ (binary, 0 or 1) and the predicted scores $\hat{y}$, as shown in Equation \ref{4}.

\begin{equation}
    \mathcal{L} = -\sum_{i=1}^{N} y_{i} \log(\hat{y}_{i})
    \label{4}
\end{equation}

\begin{figure}
    \centering
    \includegraphics[width=0.466\textwidth]{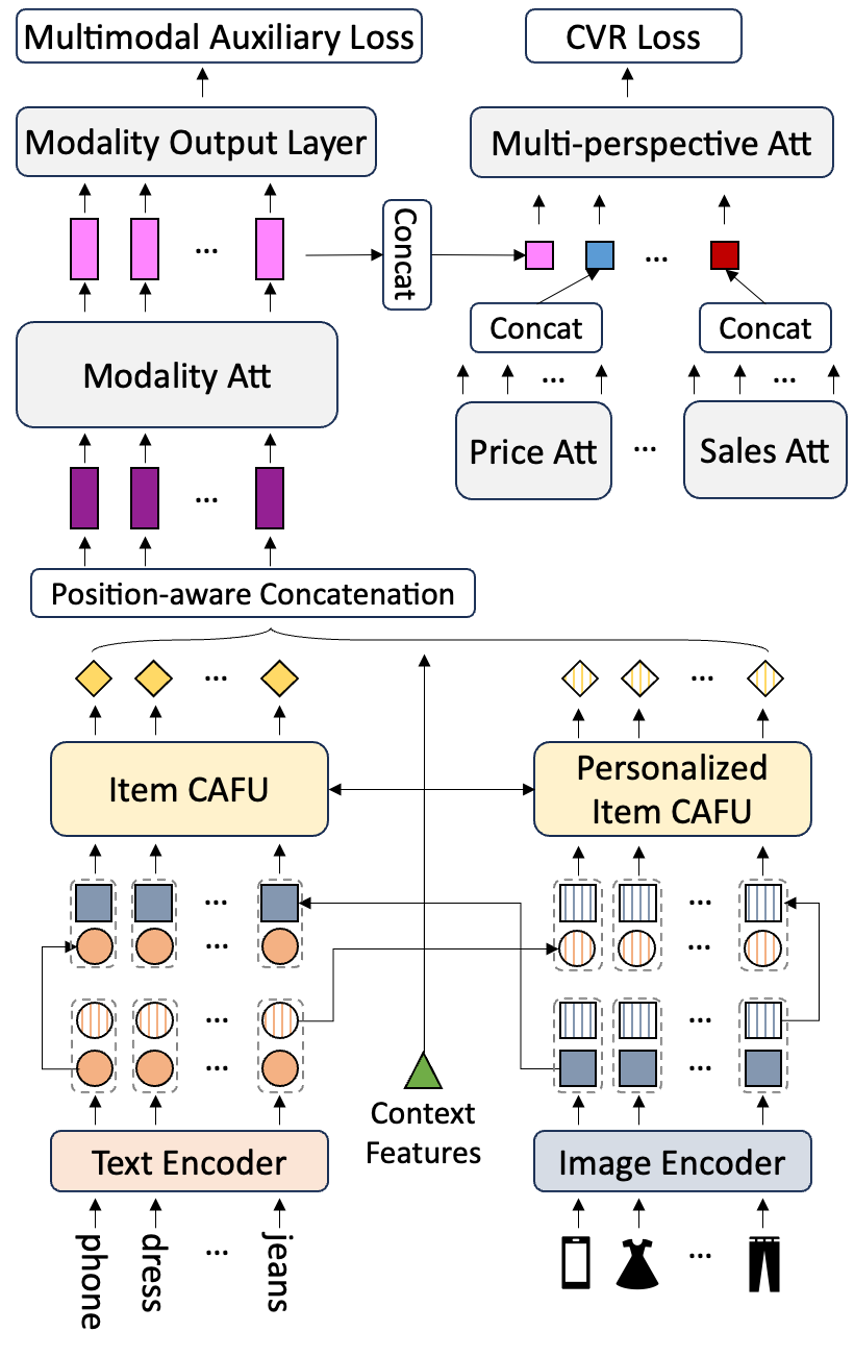}
    \caption{The framework of Advancing Re-Ranking with Multimodal Fusion and Target-Oriented Auxiliary Tasks (ARMMT).}
    \label{fig:framework}
\end{figure}

\section{Method}
% 一段介绍整体的framework，包括三个阶段，分别对应下面的四个小节
In this section, we present the overall architecture of our proposed ARMMT framework, as illustrated in Figure \ref{fig:framework}. We will explore its key components: the generation of multimodal representations for items and personalized items (derived from the interaction between items and user historical interests), the hierarchical multimodal fusion process using the Context-Aware Fusion Unit (CAFU), the integration of multimodal information with other domains through Multi-Perspective Self-Attention, and the critical multimodal auxiliary losses that enhance our model's performance.

\subsection{Multimodal representations} 
\begin{figure}
    \centering
    \includegraphics[width=0.466\textwidth]{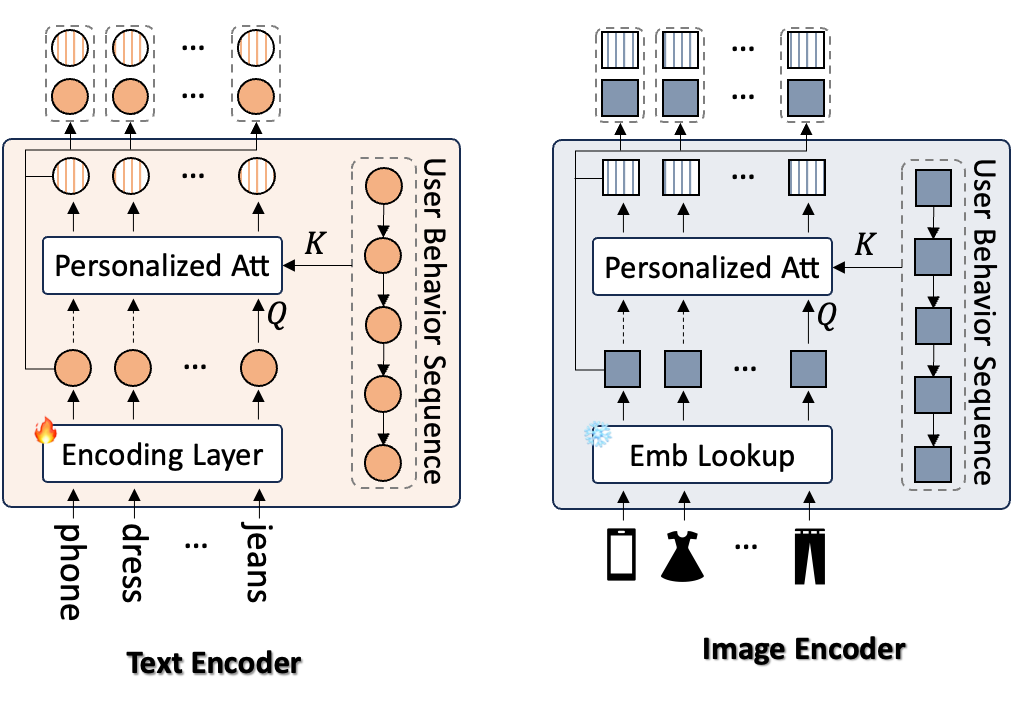}
    \caption{The encoding process of textual and image information. Effective information from user behavior sequences is extracted through multi-head attention.}
    \label{fig:encoder}
\end{figure}

% 对应图中Position-aware Concatenation之前的所有模块
% - Text Encoder
% - Image Encoder
% - Item Gate
% - Personalized Item Gate
Figure \ref{fig:encoder} illustrates the encoding processes for textual and visual features in our method, with the left and right parts depicting these processes, respectively. In the context of our sequence analysis, we derive pre-trained embeddings for items by indexing the vocabulary. Subsequently, we employ the target item as a query to apply target attention mechanisms, which effectively model the interactions between the target item and the other items in the sequence, culminating in the derivation of the final feature embeddings.

\subsubsection{Multimodal Representation of Item}
To optimize training efficiency and avoid increasing online serving pressure, we adopted a two-stage integration strategy for multimodal representations. In contrast, previous methods often fuse image and text modalities too early during the pre-training stage, which can lead to inconsistencies with user preferences and suboptimal search results. For example, when users search for dresses, they may prioritize different attributes such as color or length, which early fusion methods struggle to dynamically adjust to.

To address this, we first obtain separate representations for each modality. Specifically, for the image modality, we use frozen pre-trained image embeddings, $I_\text{img}$, extracted by fine-tuning the RegNet \cite{xu2022regnet} convolutional network on JD.com product images after object detection with YOLO4 \cite{bochkovskiy2020yolov4}. For the text modality, we use the title features directly, representing them as $I_\text{text}$. This separation allows us to maintain the integrity of each modality's features before they are combined in the ranking model.

\subsubsection{Multimodal Representation of Personalized Item}
We implemented a two-stage behavioral sequence modeling approach to accurately capture the influence of user preferences on purchasing decisions, particularly within similar product categories. For instance, a user's preference for the appearance of clothing significantly influences their decisions when purchasing similar apparel but has minimal impact on unrelated categories, such as computers. In the first stage, we use the category of the query, predicted by a large language model (LLM), to filter the user's historical behavior sequence $H$, retaining only those behaviors that match the query's category, resulting in a relevant sub-sequence $H^{\ast}$. In the second stage, we use a target attention mechanism to model interactions between these behaviors and the target item.

To capture the fine-grained features of both modalities, we model personalized item representations separately for text and images. For the image modality, we first leverage the image representations of items and those from the user's historical behaviors. Building on this foundation, we incorporate user features such as age and gender to enhance generalization across similar user groups. Furthermore, each historical behavior is detailed, including the type of interaction (e.g., click, order), frequency of actions, and the recency of the actions. These features are integral in calculating the attention weights, which are then used to create personalized item image representations $P_\text{img}$, thereby revealing the user's latent preferences in decision-making. As illustrated in Figure 2 (right side), the process of generating personalized product image representations involves obtaining image features through embedding lookup and integrating these features with the user's behavioral sequence using a personalized attention mechanism.

Similarly, for the text modality, we begin by extracting text representations from item titles, attributes, and other textual features using an encoding layer. These representations are generated for both the candidate items and the items in the user's historical behavior sequence. To further personalize these text representations $P_\text{text}$, we incorporate user demographics such as age and gender, as well as detailed historical behavior information. This includes the type of interactions (e.g., clicks, orders), the frequency of these actions, and their recency. This information is crucial for calculating the attention weights of items within the user's behavioral sequence. Figure 2 (left side) illustrates this process, highlighting how textual features are encoded and integrated with the user's behavioral sequence through a personalized attention mechanism.

\subsection{Hierarchical Multimodal Fusion}
% 内部fusion
% 外部fusion

\begin{figure}
    \centering
    \includegraphics[width=0.466\textwidth]{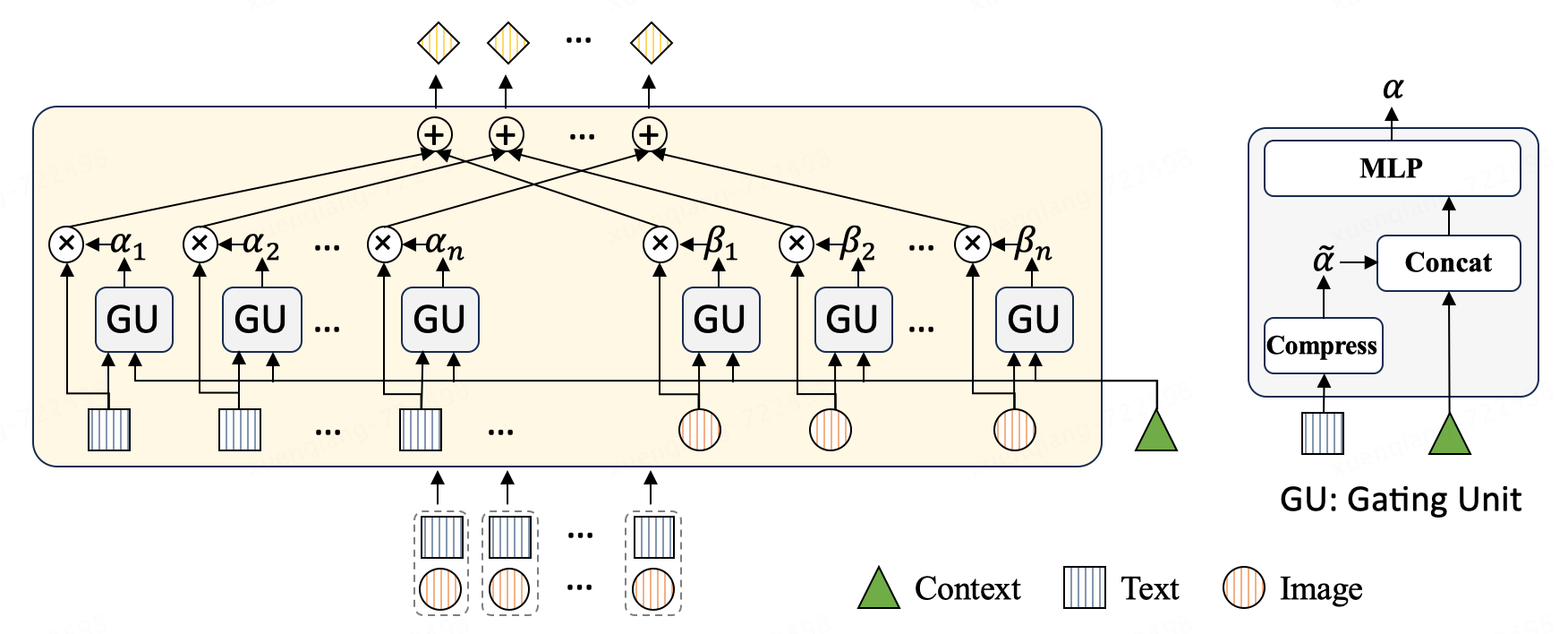}
    \caption{The diagram of the Context-Aware Fusion UNIT. In this diagram, triangles, rectangles, and circles represent context, text, and image features, respectively.}
    \Description{}
    \label{fig:cfau}
\end{figure}

\subsubsection{Context-Aware Fusion UNIT}
We introduce a Context-Aware Fusion Unit (CAFU) designed to seamlessly integrate the representations of items and personalized items from both image and text modalities. The input to this module consists of a list of features $X \in \mathbb{R}^{N \times D \times M}$, along with contextual information $C \in \mathbb{R}^{J}$, such as query features and user features, where $N$ represents the number of item candidates, $D$ denotes the item feature dimension, $M$ indicates the number of modalities (e.g., image and text), and $J$ indicates the contextual feature dimension.

Initially, we compress the representations using mean pooling to obtain an initial weighting vector $z_{n} = \left[ z_{n1}, z_{n2}, \ldots, z_{nM} \right] \in \mathbb{R}^{M}$ for each item $x_{n} \in \mathbb{R}^{D \times M}$, as shown in Equation \ref{eq:eq6}.

\begin{equation}
    z_{nm} = \frac{1}{D} \sum_{d=1}^{D} x_{ndm}
    \label{eq:eq6}
\end{equation}

Next, we concatenate the initial weights $z_{n}$ with the contextual information $C$, thereby incorporating the context into our weighting scheme, as shown in Equation \ref{eq:eq7}.

\begin{equation}
    z_{n}^{c} = \text{concat}(z_{n}, C) \in \mathbb{R}^{M + J}
    \label{eq:eq7}
\end{equation}

To calculate the weights for each modal representation, we employ a Multi-Layer Perceptron (MLP). Specifically, the weights $s_{n}$ are computed as shown in Equation \ref{eq:eq8}, where $\alpha_n$ and $\beta_n$ in Figure \ref{fig:cfau} represent the weights for different modalities to facilitate understanding.

\begin{equation}
    s_{n} = \text{softmax} \left( W_{2} \delta (W_{1} z_{n}^{c}) \right) \in \mathbb{R}^{M}
    \label{eq:eq8}
\end{equation}
where $\delta$ denotes the ReLU activation function, $W_1 \in \mathbb{R}^{\frac{M + J}{r} \times (M + J)}$ and $W_2 \in \mathbb{R}^{M \times \frac{M + J}{r}}$ are the weight matrices, and $r$ is the reduction ratio that controls the size of the hidden layer.

In the final stage, these weights are multiplied by their corresponding representations, and a sum pooling operation is performed to obtain the fused representation $B \in \mathbb{R}^{N \times D}$, which serves as the output of this module, as shown in Equation \ref{eq9} and illustrated in Figure \ref{fig:cfau}.

\begin{equation}
    b_{n} = \sum_{m=1}^{M} s_{nm} \cdot x_{nm}
    \label{eq9}
\end{equation}

This method offers a degree of interpretability, allowing for easy extraction and visual analysis of the fusion weights between different modalities. Moreover, this approach demonstrates a high degree of robustness to missing values. Utilizing this methodology, we successfully fuse the image and text representations of items into a cohesive multimodal item representation $I_{\text{mo}}$, as shown in Equation \ref{eq10}.

\begin{equation}
    I_{\text{mo}} = \text{CAFU}([I_{\text{img}}, I_{\text{text}}], C)
    \label{eq10}
\end{equation}

Similarly, a personalized multimodal item representation $P_{\text{mo}}$ is derived through the same process, as shown in Equation \ref{eq11}.

\begin{equation}
    P_{\text{mo}} = \text{CAFU}([P_{\text{img}}, P_{\text{text}}], C)
    \label{eq11}
\end{equation}

\subsubsection{Multi-Perspective Self-Attention}
Building on the integrated representations achieved through the CAFU, we apply a Multi-Perspective Self-Attention mechanism to deeply merge the multimodal field with other critical fields, such as price and sales, thereby achieving a comprehensive global feature fusion. Specifically, we combine the multimodal representations of items with personalized multimodal item representations and contextual information through concatenation, which is then processed via an MLP to obtain a unified multimodal representation, denoted as \(M_{\text{mo}}\), as shown in Equation \ref{eq:M_mo}.

\begin{equation}
    M_{\text{mo}} = \text{MLP}(\text{concat}(I_{\text{mo}}, P_{\text{mo}}, C)) 
    \label{eq:M_mo}
\end{equation}

In conventional approaches, multimodal representations \(M_{\text{mo}}\) are typically incorporated directly into the input layer, where they are combined with other features through concatenation. However, in re-ranking models, there is a risk that during self-attention computations, the model may favor strong fields, thereby overshadowing the multimodal information. To address this, we explicitly model the context by considering the interactions and variations among the multimodal information of items to be ranked. Specifically, we employ a transformer encoder architecture to capture the interactive representation, resulting in an enriched \(A_{\text{mo}}\), as shown in Equation \ref{eq:A_mo}.

\begin{equation}
    A_{\text{mo}} = \text{Attention}(Q_{\text{mo}}, K_{\text{mo}}, V_{\text{mo}}) 
    \label{eq:A_mo}
\end{equation}
where $Q_{\text{mo}}$, $K_{\text{mo}}$, and $V_{\text{mo}}$ represent the query, key, and value, respectively, all of which are linearly transformed from $M_{\text{mo}}$.

We observe a similar pattern in the main task, where users vary in their sensitivity to details such as price, promotions, sales, and quality. To address this, we process each field individually through self-attention mechanisms, yielding representations such as $A_{\text{price}}$ and $A_{\text{sales}}$. We then merge these fields with the multimodal field, resulting in a combined representation denoted as $M_{\text{main}}$, as shown in Equation \ref{eq:M_main}.

\begin{equation}
    M_{\text{main}} = \text{MLP}(\text{concat}(A_{\text{mo}}, A_{\text{price}}, A_{\text{sales}})) 
    \label{eq:M_main}
\end{equation}

Subsequently, we apply global attention on the combined representation to extract higher-order semantic features. This process helps to better align with user behavior patterns, as shown in Equation \ref{eq:A_main}.

\begin{equation}
    A_{\text{main}} = \text{Attention}(Q_{\text{main}}, K_{\text{main}}, V_{\text{main}}) 
    \label{eq:A_main}
\end{equation}
where, $Q_{\text{main}}$, $K_{\text{main}}$, and $V_{\text{main}}$ represent the query, key, and value, respectively, all of which are linearly transformed from $M_{\text{main}}$. 

Finally, we score each item in the candidate list using the softmax function to obtain the model's predicted scores $\hat{y}$, as shown in Equation \ref{eq:y_hat}.

\begin{equation}
    \hat{y} = \text{softmax}(\text{MLP}(A_{\text{main}})) 
    \label{eq:y_hat}
\end{equation}

As illustrated in Figure \ref{fig:framework}, the price attention ($A_{\text{price}}$), sales attention ($A_{\text{sales}}$), and multimodal attention ($A_{\text{mo}}$) are concatenated and then processed through Multi-Perspective Self-Attention to produce the final representation, which is used to compute the CVR loss.

\subsection{Multimodal Auxiliary Tasks}
In our study, we use supervised learning specifically designed for multimodal auxiliary tasks due to the significant impact of multimodal information on user click behavior. We incorporate click labels $y^{ctr}$ as direct user feedback. Each item in the candidate list is scored using a sigmoid function to predict the click probability $\hat{y}^{ctr}_{i}$, as shown in Equation~\ref{eq:ctr_pred}.

\begin{equation}
    \hat{y}^{ctr} = \text{sigmoid}(\text{MLP}(A_{mo}))
    \label{eq:ctr_pred}
\end{equation}

The auxiliary task loss $\mathcal{L}_{aux}$ is computed using cross-entropy, as shown in Equation~\ref{eq:aux_loss}.

\begin{equation}
    \mathcal{L}_{aux} = -\frac{1}{N} \sum_{i=1}^{N} y^{ctr}_{i} \log \hat{y}^{ctr}_{i} + (1 - y^{ctr}_{i}) \log (1 - \hat{y}^{ctr}_{i})
    \label{eq:aux_loss}
\end{equation}

\subsection{Model Training}
The main task loss $\mathcal{L}_{main}$, which is used to compute the CVR loss, is the cross-entropy between conversion labels $y$ and predicted scores $\hat{y}$ (Equation~\ref{eq:y_hat}), as shown in Equation~\ref{eq:main_loss}.

\begin{equation}
    \mathcal{L}_{main} = -\sum_{i=1}^{N} y_{i} \log \hat{y}_{i}
    \label{eq:main_loss}
\end{equation}

The final optimization objective combines both the main task loss and the auxiliary task loss, with $\lambda$ as a hyper-parameter, as shown in Equation~\ref{eq:total_loss}.

\begin{equation}
    \mathcal{L} = \mathcal{L}_{main} + \lambda \mathcal{L}_{aux}
    \label{eq:total_loss}
\end{equation}

\section{EXPERIMENTS}
\subsection{Dataset and Evaluation Metric}
\label{sec:data}
In this paper, we conduct our experiments on an internal dataset provided by JD.com. This dataset is extracted from user behavioral logs specifically related to search activities and includes only those user sessions that resulted in a conversion. The dataset is organized by user search sessions, with each row representing a different session. Each session contains 30 item samples along with their corresponding labels indicating clicks and conversions. For the training set, data was collected over a consecutive 21-day period, resulting in a substantial dataset comprising hundreds of millions of sessions and billions of products. The data from the 22nd day was reserved for the test set, allowing us to evaluate the model's performance on unseen data.

For offline evaluation, we predict the likelihood of a user's conversion for each item and employ the widely used AUC (Area Under the ROC Curve) as our performance metric. This metric is well-established in search systems, advertising, and recommendation systems, as supported by the references in \cite{cheng2016wide,mo2015image}, and represents the cumulative area under the characteristic curve, effectively quantifying the predictive accuracy of our model.

\subsection{Experimental Settings }
\subsubsection{Baseline Models}
To ensure a fair comparison, we selected some of the most representative models from the industry, such as PRM \cite{pei2019personalized} and PIER \cite{shi2023pier}, as benchmarks for our model. Our model is an optimization directly based on the PIER model. Specifically:
\begin{itemize}

\item \underline{PRM}: This stands for the context-sensitive sequence awareness model, which is a commonly used re-ranking model in the industry.
\item \underline{PIER}: This model utilizes a two-stage re-ranking architecture comprising a Sequence Generator and a Sequence Evaluator. It serves as our baseline model that has been deployed online.

\end{itemize}

\subsubsection{Implementation Details}
All our experiments, including ARMMT and the comparative baseline experiments, were implemented using TensorFlow 1.15 and run on multiple NVIDIA V100 GPUs with CUDA 10. All experiments were conducted using the same offline dataset introduced in Section \ref{sec:data}.

For the ARMMT method, we used the AdaGrad optimizer with a learning rate of 0.07 and trained the model for 20 epochs with a batch size of 128. The pre-trained ID and image vocabulary contained 8 million items, with an embedding dimension set to 32. The Transformer encoder consisted of two layers, each employing multi-head self-attention with 6 heads, and each head having a dimension of 32. The weight of the auxiliary loss, $\lambda$, was set to 1. This configuration was chosen to facilitate efficient convergence and ensure effective learning from the training data.

\subsection{Performance Comparison}
In this subsection, we conduct a comparative study of our proposed ARMMT model, which is benchmarked against PIER \cite{shi2023pier} and PRM \cite{pei2019personalized}, as presented in Table \ref{tab:base}. PIER, characterized by its two-stage architecture and the introduction of an evaluator for scoring sequence generation, shows a notable enhancement in the AUC metric compared to the traditional PRM. Extending PIER, our ARMMT model incorporates visual cues into the re-ranking process, achieving alignment between the modal information of the model's inputs and the user's browsing behavior. This enhancement yields the highest performance, with an offline AUC score of 0.9647, marking a 0.0005 increase over PIER. Given that the re-ranking module focuses on scoring only the top 30 items from the ranking model, this AUC improvement of 0.0005 is statistically meaningful, thereby validating the efficacy of our ARMMT model.

% In this subsection, we conduct a comparative study of our proposed ARMMT method, which is benchmarked against the PIER method by Shi et al.\cite{shi2023pier} and the PRM method by Pei et al.\cite{pei2019personalized}, as presented in Table \ref{tab:base}. The PIER method\cite{shi2023pier}, characterized by its two-stage architecture and the introduction of an evaluator for scoring sequence generation, shows a notable enhancement in the AUC metric compared to the traditional PRM method \cite{pei2019personalized}. Extending the PIER method \cite{shi2023pier}, our ARMMT incorporates visual cues into the re-ranking process, achieving alignment between the modal information of the model's inputs and the user's browsing behavior. This enhancement yields the highest performance, with an offline AUC score of 0.9647, marking a 0.0005 increase over the PIER method. Given that the re-ranking module focuses on scoring only the top 30 items from the ranking model, this AUC improvement of 0.0005 is statistically meaningful, thereby validating the efficacy of our ARMMT method.

\begin{table}[htbp]
  \centering
  \caption{Performance comparison between the baseline models and our proposed ARMMT method. Bold indicates the best result, underlined indicates the second-best, and the AUC gain shows the improvement achieved by ARMMT over the second-best result.}
    \begin{tabular}{ccc}
    \toprule
    Methods & AUC & AUC Gain\\
    \midrule
    PRM \cite{pei2019personalized} & 0.9636 & -\\
    PIER \cite{shi2023pier} & \underline{0.9642} & -\\
    \midrule
    ARMMT & \textbf{0.9647} & \textbf{0.0005}\\
    \bottomrule
    \end{tabular}%
  \label{tab:base}%
  \vspace{0.3cm}
\end{table}%

\subsection{Ablation Studies}
In this subsection, we perform an extensive ablation analysis to quantitatively assess the impact of various architectural components on the model's performance. The analysis includes the following experimental setups:
(1) \textbf{PIER \cite{shi2023pier}}: A benchmark model relying solely on textual and ID-embedding features, excluding image embeddings.
(2) \textbf{w/o CAFU and Auxiliary Tasks}: A model that directly incorporates image embeddings, bypassing the Context-Aware Fusion Unit and auxiliary losses.
(3) \textbf{w/o Auxiliary Tasks}: A model using CAFU for multimodal fusion, excluding auxiliary losses.
(4) \textbf{ARMMT}: The complete framework integrating both CAFU and auxiliary tasks.

% In this subsection, we perform an extensive ablation analysis to quantitatively assess the impact of various architectural components on the model's performance. The analysis includes the following experimental setups: (1) A benchmark PIER \cite{shi2023pier} model that relies solely on textual and ID-embedding features, excluding the integration of item-specific image data. (2) An intermediary model that incorporates multimodal embeddings directly, bypassing the Context-Aware Fusion UNIT and the corresponding auxiliary losses, to evaluate the inherent benefits of multimodal integration, denoted as w/o ARMMT. (3) An enhanced model that employs our proposed Context-Aware Fusion UNIT for the fusion of multimodal representations, while excluding task-specific auxiliary losses, to isolate the contribution of the CAFU module to the refinement of item representations, denoted as w/o Auxiliary Tasks. (4) The complete ARMMT framework, integrating both the Context-Aware Feature Fusion Unit and Multimodal Auxiliary Tasks, to exemplify the synergistic effects of our comprehensive approach.

\begin{table}[htbp]
  \centering
  \caption{Ablation study of ARMMT.}
    \begin{tabular}{ccc}
    \toprule
    Methods & AUC & AUC gain\\
    \midrule
    PIER \cite{shi2023pier}  & 0.9642 & -\\
    w/o CAFU and Auxiliary Tasks & 0.9643 & 0.0001\\
    w/o Auxiliary Tasks & 0.9645 & 0.0003\\
    \midrule
    ARMMT & \textbf{0.9647} & \textbf{0.0005}\\
    \bottomrule
    \end{tabular}%
  \label{tab:ablation}%
  \vspace{0.6cm}
\end{table}%

As shown in Table \ref{tab:ablation}, our ablation experiments on the PIER \cite{shi2023pier} baseline model yielded varying degrees of improvement in the AUC metric. Directly incorporating image embeddings into the re-ranking model resulted in a 0.0001 increase in AUC compared to PIER \cite{shi2023pier}, validating that incorporating image embeddings can enhance the model's predictive accuracy. The introduction of CAFU and auxiliary tasks brought about additional AUC improvements of 0.0002 and 0.0002, respectively. These results suggest that the strategic design of CAFU and the incorporation of auxiliary tasks effectively enhance multimodal feature fusion, thereby improving prediction precision. Our comprehensive ARMMT model achieves the best performance, with an offline AUC of 0.9647.

% As demonstrated in Table \ref{tab:ablation}, our ablation experiments on the PIER \cite{shi2023pier} baseline model have yielded varying degrees of improvement in the AUC metric. Specifically, integrating image information directly into the re-ranking model resulted in a 0.0001 increase in AUC compared to PIER \cite{shi2023pier}, validating that incorporating image information can better simulate the input signals received during the user decision-making process, thereby enhancing the model's predictive accuracy. Furthermore, the introduction of the multimodal feature fusion module (CAFU) and auxiliary tasks brought about additional AUC improvements of 0.0003 and 0.0005, respectively. These results suggest that the strategic design of the CAFU module and the incorporation of auxiliary tasks effectively enhance the efficacy of multimodal feature fusion, thereby further improving the precision of predictions. Incorporating these improvements, our comprehensive ARMMT method achieves the best performance metrics in offline evaluations, with an offline AUC of 0.9647.

\subsection{Online Results}
On JD.com's platform, which boasts tens of millions of daily active users, we conducted a 7-day A/B test, allocating 10\% of the user traffic to each experimental variant. The control variant (Variant A) used the PIER \cite{shi2023pier} model, whereas the experimental variant (Variant B) implemented the novel ARMMT model proposed in our research.

As indicated in Table \ref{tab:abtest}, the ARMMT model demonstrated superior performance across all pivotal metrics. Notably, it yielded a statistically significant 0.22\% rise in CVR (p-value = 0.04) \cite{xu2024optimizing}, alongside a 0.03\% uplift in CTR and a substantial 0.49\% increase in GMV. These improvements have led to a marked escalation in business revenue. The experimental results are consistent with our prior offline analyses, confirming the ARMMT model's effectiveness in integrating visual information into the re-ranking process. This seamless integration significantly improves sorting efficiency, ensures a more precise alignment with user preferences, and notably enhances the platform's conversion rate.

In light of these positive outcomes, the ARMMT model was deployed on the platform in early 2024, providing substantial benefits to hundreds of millions of users.

% As indicated in Table \ref{tab:abtest}, the ARMMT model demonstrated superior performance across all pivotal metrics. It yielded a 0.03\% uplift in the Click-Through Rate (CTR), a statistically significant 0.22\% rise in the Conversion Rate (CVR) \cite{xu2024optimizing}, and a substantial 0.49\% increase in the Gross Merchandise Volume (GMV). The improvements in these metrics have led to a marked escalation in business revenue. The experimental results are in alignment with our earlier offline analyses, affirming the ARMMT model's proficiency in integrating visual information into the re-ranking process. This seamless integration significantly elevates sorting efficiency, ensures a more nuanced alignment with user preferences, and markedly strengthens the platform's conversion rate.

\begin{table}[htbp]
  \centering
  \caption{Online A/B test of ARMMT. The improvements are averaged over 7 days in April 2024.}
    \begin{tabular}{ccc}
    \toprule
    Online Metrics & Improvement & p-value \\
    \midrule
    CVR  & \textbf{0.22\%} & \textbf{0.043} \\
    GMV   & 0.49\% & 0.121 \\
    CTR  & 0.03\% & 0.668 \\
    \bottomrule
    \end{tabular}%
  \label{tab:abtest}%
  % \vspace{0.6cm}
\end{table}%

\section{conclusion}
This study introduces an innovative attention-based multimodal fusion method for re-ranking in e-commerce search. By integrating textual and visual information, it overcomes the limitations of traditional re-ranking models that rely solely on unimodal data, providing a more intuitive representation of the content users interact with. Additionally, an auxiliary task designed to predict click-through rates aligns closely with the ranking task, ensuring the effective utilization of multimodal information. Empirical results demonstrate that these combined approaches enhance user satisfaction and conversion rates, highlighting significant practical implications for the e-commerce industry and providing a competitive edge in the market. Furthermore, this research opens new avenues for incorporating additional modalities and dynamic ranking objectives, advancing re-ranking technologies and offering practical guidance for e-commerce platforms to enhance personalized shopping experiences.


\bibliographystyle{ACM-Reference-Format}
\balance
\bibliography{sample-base}

% \section*{Company Portrait}
% JD.com, Inc., also known as Jingdong, is a Chinese e-commerce company headquartered in Beijing. It is one of the two massive B2C online retailers in China by transaction volume and revenue, a member of the Fortune Global 500. When classified as a tech company, it is the largest in China by revenue and 7th in the world in 2021.

% \section*{Presenter profiles}
% \noindent\textbf{Enqiang Xu} is a researcher in the Department of Search and Recommendation at JD.com Beijing. 
% He received his master degree in School of Mathematical Sciences, Peking University.
% His research focuses on information retrieval and natural language processing.

\end{document}